\documentclass[amsfonts,nofootinbib,prd,aps]{revtex4}
\usepackage{graphicx}

\begin{document}

\title{Quantum scalar field in quantum gravity:\\ the propagator and
  Lorentz invariance in the spherically symmetric case}

\author{ Rodolfo Gambini$^{1}$,
Jorge Pullin$^{2}$,
Saeed Rastgoo$^{1}$}
\affiliation {
1. Instituto de F\'{\i}sica, Facultad de Ciencias,
Igu\'a 4225, esq. Mataojo, Montevideo, Uruguay. \\
2. Department of Physics and Astronomy, Louisiana State University,
Baton Rouge, LA 70803-4001}

\begin{abstract}
  We recently studied gravity coupled to a scalar field in spherical
  symmetry using loop quantum gravity techniques. Since there are
  local degrees of freedom one faces the ``problem of dynamics''.
  We attack it using the ``uniform discretization technique''. We
  find the quantum state that minimizes the value of the master
  constraint for the case of weak fields and curvatures. The state has
  the form of a direct product of Gaussians for the gravitational
  variables times a modified Fock state for the scalar field. In this
  paper we do three things. First, we verify that the previous state
  also yields a small value of the master constraint when one
  polymerizes the scalar field in addition to the gravitational
  variables. We then study the propagators for the polymerized scalar
  field in flat space-time using the previously considered ground
  state in the low energy limit. We discuss the issue of the Lorentz
  invariance of the whole approach. We note that if one uses real
  clocks to describe the system, Lorentz invariance violations are
  small.  We discuss the implications of these results in the light of
  Ho\v{r}ava's {\em Gravity at the Lifshitz point} and of the argument
  about potential large Lorentz violations in interacting field
  theories of Collins {\em et. al.}
\newline
\\
This work is dedicated to Josh Goldberg for his many contributions to our
understanding of space-time.
\end{abstract}

\maketitle

\section{Introduction}
In previous work \cite{nosotros} (hereafter referred to as Paper I),
we have studied a spherically symmetric scalar field coupled to
spherically symmetric gravity in the loop representation. Using
symmetry adapted variables, one is left with a diffeomorphism and
Hamiltonian constraint that have a non-trivial constraint algebra with
structure functions, pretty much like in the full theory. We gauge
fixed the diffeomorphism constraint to make things simpler but the
remaining Hamiltonian constraint has an algebra with itself still
involving structure functions and, up to present, no one has found a
gauge fixing that would avoid this problem that leads to a Hamiltonian
that is local \cite{husain}. Having structure functions poses problems
for the Dirac quantization procedure. We decided to handle the
situation using the ``uniform discretization'' approach
\cite{uniform}. In that approach one discretizes the theory in such a
way that the evolution equations are generated by the {\em master
  constraint} (the sum of the squares of the constraints)
\cite{master}. One studies the spectrum of the quantum master
constraint. If zero is in the spectrum the associated eigenstate
corresponds to the continuum limit. If zero is not in the spectrum,
one is left with a quantum theory that has a fundamental level of
discreteness, but that can approximate the continuum theory well in
many circumstances of interest.  We could not conclusively prove that
zero was in the kernel in the model studied. We proceeded
variationally by proposing a trial state which depended on a set of
parameters and minimized the value of the master constraint as a
function of those parameters.  The resulting state had a very small
value of the master constraint for lattice spacings that are large
compared to the Planck scale (but very small by, say, particle physics
scales) and therefore approximates continuum general relativity in
large scales very well. For simplicity we used a ``polymer''
representation for the gravitational variables but followed a regular
quantization for the scalar field.  The state has the form of a direct
product of Gaussians for the gravitational variables at each lattice
site times a modified Fock vacuum for the scalar field variables (the
modification is due to the fact that the background is not globally
flat, in 1+1 dimensions the zero point energy of the vacuum generates
a deficit angle, and also that we incorporate quantum corrections to
the background geometry).

In this paper we will do three things. First we will verify that the
vacuum state we just discussed is a good vacuum for the polymerized
theory, at least in the case in which the polymerization parameter is
small. We will compute the expectation value of the master constraint
for the fully polymerized theory in the vacuum state to leading order
in the polymerization parameter and show that the resulting terms are
very small. Second, we will study the low energy propagator for the
scalar field on the above discussed quantum state. We will see that
one has different options for polymerizing the scalar field and this
will lead to different types of propagators. Generically they fall
within the class of propagators considered by Ho\v{r}ava
\cite{horava}. We will again work in the limit in which the
polymerization parameter is small. The resulting propagators are not
Lorentz invariant.  We will analyze these effects in the light of the
work of Collins, P\'erez, Sudarsky, Urrutia and Vucetich
\cite{collins} that shows that even Lorentz violations of Planck scale
can have catastrophic effects when one considers interacting quantum
field theories. We will see that their argument does not apply to this
model if one considers real clocks to parameterize time.

It is worthwhile mentioning related recent work. Husain and
Kreienbuehl \cite{Husain:2010gb} consider the polymerization of a
scalar field without assuming spherical symmetry and proceed to define
creation and annihilation operators for the polymerized theory.
More recently, Hossain,
Husein and Seahra \cite{HoHuSe} have analyzed the propagator in that
context and have found Lorentz violations.  Their work cannot be directly
compared to ours for reasons we will discuss in section
\ref{field}.  Laddha and Varadarajan \cite{Laddha:2010hp}
consider a scalar field in $1+1$ dimensions but parameterized,
including the embedding variables in their treatment. They are
apparently able to recover Lorentz invariance exactly so the
connection to our work is at the moment unclear.

The plan of the paper is as follows: in the next section we will
discuss how justified one is in using the Fock vacuum for the scalar
field in the context of a polymerized theory.  In section III we
discuss the polymerization of the field and the polymerization of the
canonically conjugate momentum of the field and compute the resulting
propagators. In section IV we address the issue of Lorentz invariance.
We end the paper with a discussion.

\section{Appropriateness of using the Fock vacuum for the scalar field}

In Paper I we minimized the master constraint using a variational
technique that used as trial state one that consisted of Gaussians
centered around flat space-time for the gravitational variables times
a (curved space, quantum corrected) Fock vacuum for the scalar
field. The use of the Fock vacuum appeared compelling in part due
to the fact that we were not polymerizing the scalar field in our
treatment. Since in this paper we will be polymerizing the scalar
field, it begs the question of the appropriateness of continuing
to use the Fock vacuum. In this section we would like to show
that the Fock vacuum still yields a very small value for the master
constraint even if one polymerizes the scalar field variables.

We start by considering the Hamiltonian of gravity coupled to a
scalar field in spherical symmetry we considered in Paper I,
\begin{equation}\label{ham}
H= H_{\rm vac}+2 G\,H_{\rm matt},
\end{equation}
where
\begin{eqnarray}
H_{\rm vac}&=&\left( -x -x K_\varphi^2+\frac{x^3}
{(E^\varphi)^2}\right)'=\partial H_v(x)/\partial x,\\
H_{\rm matt} &=& \frac{P_\phi^2}{2(E^\varphi)^2}
+\frac{x^4 (\phi')^2}{2(E^\varphi)^2} - \frac{x K_\varphi P_\phi
\phi'}{E^\varphi}.
\end{eqnarray}
We will now rescale the variables,
\begin{eqnarray}
P_\phi^{\rm orig} = x P_\phi^{\rm new}, \label{rescal1}\\
\phi^{\rm orig}=\phi^{\rm new}/x, \label{rescal2}
\end{eqnarray}
and will drop the ``new'' superscript from now on to economize
in the notation. The matter Hamiltonian then becomes,
\begin{equation}
H_{\rm matt} = \frac{H^{(1)}}{\left(E^\varphi\right)^2}
+\frac{ H^{(2)} K_\varphi}{E^\varphi},
\end{equation}
where
\begin{eqnarray}
H^{(1)}&=&\frac{1}{2} P_\phi^2 x^2 +\frac{1}{2} \phi^2 -x\phi' \phi
+\frac{1}{2} x^2 (\phi')^2,
\\
H^{(2)}&=&\phi P_\phi-x \phi' P_\phi.
\end{eqnarray}
We now proceed to discretize and polymerize the matter Hamiltonian,
\begin{equation}
H_{\rm matt}(i) = \frac{H^{(1)}(i)}{\left(E^\varphi(i)\right)^2}
+\frac{ H^{(2)}(i) \sin\left(\rho K_\varphi(i)\right)}{\rho
  E^\varphi(i)},
\label{matterH}
\end{equation}
where,
\begin{eqnarray}
H^{(1)}(i)&=&
\frac{\epsilon}{2}P_\phi^2(i) x(i)^2
+\frac{\epsilon^3 \sin^2\left(\beta \phi(i)\right)}{2\beta^2}
- \frac{\epsilon^2 x(i)}{\beta^2} \sin\left(\beta \phi(i)\right)
\sin\left(\beta\left(\phi(i+1)- \phi(i)\right)\right)\label{H1}\\
&&+ \frac{\epsilon x(i)^2}{2\beta^2}
\sin^2\left(\beta\left(\phi(i+1)- \phi(i)\right)\right),\nonumber\\
H^{(2)}(i)&=&
\frac{P_\phi(i)}{\beta}
\left( \epsilon\sin\left(\beta \phi(i)\right)
-x(i) \sin\left(\beta \left(\phi(i+1)-\phi(i)\right)\right)\right).\label{H2}
\end{eqnarray}
We now write the complete Hamiltonian but expand the trigonometric
functions in $\beta$ and keep the two lowest orders, e.g.  $\sin(\beta
\phi)/\beta \sim \phi-\beta^2\phi^3/6$, we do this so it is clear that
to leading order one will have the same results as in Paper I, and the
next order will be the corrections introduced by the polymerization
and we can analyze their influence.  We get for (\ref{H1}) and (\ref{H2}),
\begin{eqnarray}
H^{(1)}(i)&=&H^{(1)}_{\textrm{lead}}(i)+H^{(1)}_{\textrm{corr}}(i),\\
H^{(2)}(i)&=&H^{(2)}_{\textrm{lead}}(i)+H^{(2)}_{\textrm{corr}}(i),
\end{eqnarray}
for which ``lead'' refers to leading order and ``corr'' refers to the correction terms based on the above expansion of matter Hamiltonian in $\beta$ and
\begin{eqnarray}
H^{(1)}_{\textrm{lead}}(i)&=&\frac{\epsilon}{2}P_\phi(i)^2 x(i)^2
+\frac{1}{2}\epsilon^3 \phi(i)^2
+\frac{1}{2}\epsilon x(i)^2\left(\phi(i+1)-\phi(i)\right)^2
-\epsilon^2 x(i)\phi(i)\left(\phi(i+1)-\phi(i)\right),\label{H1lead}\\
H^{(1)}_{\textrm{corr}}(i)&=&\frac{\epsilon^3\beta^2}{6}\bigg(
- x(i)^2 \frac{\left(\phi(i+1)-\phi(i)\right)^4}{\epsilon^2}
+\frac{x(i)\left(\phi(i+1)-\phi(i)\right)\phi(i)^3}{\epsilon},\label{H1corr}\\
&&+\frac{x(i)\left(\phi(i+1)-\phi(i)\right)^3\phi(i)}{\epsilon}
-\phi(i)^4\bigg)\nonumber\\
H^{(2)}_{\textrm{lead}}(i)&=&\epsilon\left(
-\frac{x(i)P_\phi(i)\left(\phi(i+1)-\phi(i)\right)}{\epsilon}
+P_\phi(i)\phi(i)
\right),\label{H2lead}\\
H^{(2)}_{\textrm{corr}}(i)&=&\frac{\epsilon\beta^2}{6}\left(
\frac{x(i) P_\phi(i)\left(\phi(i+1)-\phi(i)\right)^3}{\epsilon}
-P_\phi(i)\phi(i)^3\right).\label{H2corr}
\end{eqnarray}
These should be put into (\ref{matterH}) to give the matter
Hamiltonian expanded in $\beta$. We are now going to focus on the
master constraint. It can be written as,
\begin{equation}
{\mathbb H}(i)=
c_{11}(i) \left(H^{(1)}(i)\right)^2
+c_1(i) H^{(1)}(i)
+c_{12}(i) H^{(1)}(i)H^{(2)}(i)
+c_{22}(i) \left(H^{(2)}(i)\right)^2
+c_2(i) H^{(2)}(i)
.\label{MasConstGen}
\end{equation}
Where the $c$ coefficients depend only on the gravitational
variables.
Substituting the leading order terms of (\ref{H1lead}) and
(\ref{H2lead}) into (\ref{MasConstGen}), yields the results of paper I
(taking into account the re-scalings (\ref{rescal1}) and
(\ref{rescal2})). What we want to show now is that substituting the
correction terms into the master constraint (\ref{MasConstGen}) and
taking its expectation value with respect to the trial vacuum state of
the previous paper, yields corrective terms which are very small.

For this, we observe that the contribution of the correction terms to
the master constraint can be written as,
\begin{eqnarray} {\mathbb H}_{\textrm{corr}}(i)&=&
  c_{11}(i) \left(H^{(1)}_{\textrm{lead}}(i)
    H^{(1)}_{\textrm{corr}}(i)\right)
+c_1(i) H^{(1)}_{\textrm{corr}}(i)
+ c_{12}(i)\left( H^{(1)}_{\textrm{lead}}(i)
  H^{(2)}_{\textrm{corr}}(i)
+ H^{(2)}_{\textrm{lead}}(i) H^{(1)}_{\textrm{corr}}(i)
\right)\label{MastConstCorr}\\
  &&+c_{22}(i)\left(H^{(2)}_{\textrm{lead}}(i)
    H^{(2)}_{\textrm{corr}}(i)\right)+ c_2(i)
  H^{(2)}_{\textrm{corr}}(i)+c_{00}(i)\nonumber
\end{eqnarray}
and we want to show that $\langle \psi^{\rm trial}_{\vec{\sigma}}\vert
{\mathbb H}_{\textrm{corr}}(i) \vert \psi^{\rm
  trial}_{\vec{\sigma}}\rangle$ is very small.

Our strategy is the following: We compute the dominant terms by
first going to the continuum limit and
writing (\ref{H1lead})-(\ref{H2corr}) in their continuum limit form
by using,
\begin{eqnarray}
P_\phi(i)&=&\epsilon P_{\phi}(x),\\
E^\phi(i)&=&\epsilon E^{\phi}(x),\\
\phi(i)&=&\phi(x),\\
\frac{\phi(i+1)-\phi(i)}{\epsilon}&=&\frac{\partial\phi(x)}{\partial x}.
\end{eqnarray}
We then substitute in the result the continuum form of
(\ref{H1lead})-(\ref{H2corr}) that we just calculated and also
the Fourier expansions of the
$\phi(x)$ and its conjugate momentum $P_{\phi}(x)$ fields, which are,
\begin{equation}
\phi(x,t) = \frac{1}{2}\int_{-\infty}^{\infty} d\omega
\frac{\left(C(\omega) e^{-i\omega t}
+\bar{C}(\omega) e^{i\omega t}\right)  \sin(\omega x)}{\sqrt{\pi{\omega}}},
\end{equation}
and
\begin{equation}
P_{\phi}(x,t) = \frac{1}{2}\int_{-\infty}^{\infty} d\omega
\frac{-i\omega \left(C(\omega) e^{-i\omega t}
-\bar{C}(\omega) e^{i\omega t}\right)  \sin(\omega x)}{\sqrt{\pi{\omega}}}.
\end{equation}
Next, using the expanded version of the terms
(\ref{H1lead})-(\ref{H2corr}) resulting from the substitution of the
Fourier expansion of the fields, we find the individual terms that
constitute (\ref{MastConstCorr}), meaning the terms that appear
multiplied by $c_m$ and $c_{kl}$'s in (\ref{MastConstCorr})
(i.e. $H^{(1)}_{\textrm{lead}}(x) H^{(1)}_{\textrm{corr}}(x)$ etc).

{}From here on, let us focus on one of the individual terms that make up
(\ref{MastConstCorr}) (an arbitrary one). We can then repeat the
process for all the other terms. We proceed to find the portions of the
individual terms that do not have vanishing expectation values, taking
into account that $C(\omega)$ and $\bar{C}(\omega)$ are annihilation
and creation operators. It turns out that we encounter terms with four
$C(\omega)$ and/or $\bar{C}(\omega)$ operators for non-cross terms
like for example $H^{(1)}_{\textrm{corr}}(x)$ and six of them in cross
terms like $H^{(1)}_{\textrm{lead}}(x)
H^{(1)}_{\textrm{corr}}(x)$. Then the parts of the non-cross terms
with four operators with non-vanishing expectation values just include
the terms with
\begin{eqnarray}
C_4 \bar{C}_3 C_2 \bar{C}_1\label{4c1},\\
C_4 C_3 \bar{C}_2 \bar{C}_1\label{4c2},
\end{eqnarray}
where we wrote $C(\omega_{1})$ as $C_1$, etc.  for the sake of
brevity. Also the parts of the cross terms with six operators with
non-vanishing expectation values turn out to include only the terms with
\begin{eqnarray}
C_6 C_5 C_4 \bar{C}_3 \bar{C}_2 \bar{C}_1\label{6c1},\\
C_6 C_5 \bar{C}_4 C_3 \bar{C}_2 \bar{C}_1\label{6c2},\\
C_6 \bar{C}_5 C_4 C_3 \bar{C}_2 \bar{C}_1\label{6c3},\\
C_6 C_5 \bar{C}_4 \bar{C}_3 C_2 \bar{C}_1\label{6c4},\\
C_6 \bar{C}_5 C_4 \bar{C}_3 C_2 \bar{C}_1\label{6c5}.
\end{eqnarray}
Using the commutation relation
\begin{equation}
\left[\hat{C}(\omega_1),\hat{\bar{C}}(\omega_2)\right]=\delta(\omega_1-\omega_2),
\end{equation}
we evaluate the expectation values of the relevant parts (of the
individual term) we are working with. We will have
\begin{eqnarray}
\langle C_4 \bar{C}_3 C_2 \bar{C}_1 \rangle &=& \delta(\omega_4-\omega_3)\delta(\omega_2-\omega_1),\\
\langle C_4 C_3 \bar{C}_2 \bar{C}_1 \rangle &=& \delta(\omega_4-\omega_2)\delta(\omega_3-\omega_1) + \delta(\omega_4-\omega_1) \delta(\omega_3-\omega_2),\\
\langle C_6 C_5 C_4 \bar{C}_3 \bar{C}_2 \bar{C}_1 \rangle&=& \delta(\omega_6-\omega_3)[\delta(\omega_5-\omega_2)\delta(\omega_4-\omega_1) + \delta(\omega_5-\omega_1)\delta(\omega_4-\omega_2)]\\
&&+\delta(\omega_6-\omega_2)[\delta(\omega_5-\omega_3)\delta(\omega_4-\omega_1) + \delta(\omega_5-\omega_1)\delta(\omega_4-\omega_3)]\nonumber\\
&&+\delta(\omega_6-\omega_1)[\delta(\omega_5-\omega_3)\delta(\omega_4-\omega_2) + \delta(\omega_5-\omega_2)\delta(\omega_4-\omega_3)],\nonumber\\
\langle C_6 C_5 \bar{C}_4 C_3 \bar{C}_2 \bar{C}_1 \rangle&=&\delta(\omega_6-\omega_4)[\delta(\omega_5-\omega_2)\delta(\omega_3-\omega_1) + \delta(\omega_5-\omega_1)\delta(\omega_3-\omega_2)]\\
&&+\delta(\omega_5-\omega_4)[\delta(\omega_6-\omega_2)\delta(\omega_3-\omega_1) + \delta(\omega_6-\omega_1)\delta(\omega_3-\omega_2)],\nonumber\\
\langle C_6 \bar{C}_5 C_4 C_3 \bar{C}_2 \bar{C}_1 \rangle&=& \delta(\omega_6-\omega_5)[\delta(\omega_4-\omega_2)\delta(\omega_3-\omega_1) + \delta(\omega_4-\omega_1)\delta(\omega_3-\omega_2)],\\
\langle C_6 C_5 \bar{C}_4 \bar{C}_3 C_2 \bar{C}_1 \rangle&=&\delta(\omega_2-\omega_1)[\delta(\omega_6-\omega_4)\delta(\omega_5-\omega_3) + \delta(\omega_6-\omega_3)\delta(\omega_5-\omega_4)],\\
\langle C_6 \bar{C}_5 C_4 \bar{C}_3 C_2 \bar{C}_1 \rangle&=&\delta(\omega_6-\omega_5)\delta(\omega_4-\omega_3)\delta(\omega_2-\omega_1).
\end{eqnarray}

Finally, we add up the expectation values of the relevant parts
resulting from the previous step (these results are the non-vanishing
expectation-values parts of the individual term), to get the complete
expectation value of the individual term we chose. We now repeat the
procedure to get the complete expectation value of all the other individual
terms that build up (\ref{MastConstCorr}) and after that, add up all the
results to get the expectation value of ${\mathbb
  H}_{\textrm{corr}}(x)$.

Then we convert the resulting expectation value $\langle {\mathbb
  H}_{\textrm{corr}}(x) \rangle$ back to its discrete form, $\langle
{\mathbb H}_{\textrm{corr}}(i) \rangle$, by reversing the continuum
limit and neglecting highly oscillating terms like $\sin(\frac{n\pi
  x}{\epsilon})$ and the similar cosine and ${\rm Ci}$ terms. Expanding
the result in $\ell_{p}$, collecting the terms of the order of
$\beta^2$, and expanding it in $\epsilon$, we find that the
leading term of corrections are of order
\begin{equation}
  \langle {\mathbb  H}_{\textrm{corr}}(x) \rangle\sim\frac{\ell_p^5
\ln\left(\frac{\pi x}{\epsilon}\right)^2}{\epsilon \pi x^4}\beta^2.
\end{equation}
This leading term is actually the expectation value of a master
constraint density. In order to get the expectation value of the
master constraint itself, we need to integrate the above term with
respect to $x$ which will yield relevant terms of order
\begin{equation}
\int_{\epsilon}^{L} \langle {\mathbb{H}}_{\textrm{corr}}(x) \rangle dx
\sim \frac{\ell_p^5}{\epsilon^4}\beta^2.
\end{equation}
But in the previous paper, for the master constraint density, we had
the leading order result of the form
\begin{equation}
\langle {\mathbb  H}_{\textrm{lead}}(x) \rangle\sim\frac{\ell_p^3}{\epsilon x^2},
\end{equation}
where here ``lead'' means not the leading term of the corrections but
the leading term of the expectation value of the master constraint
density. Thus integrating with respect to $x$ as above will give us
the master constraint relevant terms of the order
\begin{equation}
\int_{\epsilon}^{L} \langle {\mathbb  H}_{\textrm{lead}}(x) \rangle dx\sim \frac{\ell_p^3}{\epsilon^2}.
\end{equation}
Thus we see that the corrections to the master constraint are indeed
considerably smaller than the leading contributions, provided that the
lattice spacing $\epsilon$ is large compared to the Planck length (but
still small compared to particle physics scales, as we discussed in
more detail in paper I).

\section{Low energy propagators for the scalar field}
\subsection{The standard treatment}

In the previous section we have shown that the vacuum of the theory
is well approximated by the tensor product state obtained variationally
in Paper I. For such a state the space-time metric is locally flat
with a global deficit angle. We would like to study the propagator of the
polymerized scalar field in such a metric and determine possible corrections
to the usual propagator introduced by the polymerization. We will
study the propagator perturbatively in $\beta$, the polymerization
coefficient, assuming the latter is small.

The Hamiltonian for a scalar field in spherical symmetry on a locally
flat background is given by,
\begin{equation}
H = \frac{P_\phi^2}{2x^2} + \frac{x^2 \left(\phi'\right)^2}{2}
\end{equation}
where $x$ is the radial coordinate. In order to study the modes
of the resulting equation of motion it is convenient to introduce
a rescaling $\tilde{P}_\phi=P_\phi/x$ and $\tilde{\phi}=x \phi$. We
drop the tildes from now on to simplify the notation. The
Hamiltonian then becomes (ignoring boundary terms),
\begin{equation}
H = \frac{{P}_\phi^2}{2} + \frac{\left(\phi'\right)^2}{2}.
\end{equation}
The resulting wave equation can be solved in Fourier space,
\begin{equation}
\phi(x,t)=\frac{1}{2}
\int_{-\infty}^\infty dk \frac{\left(C(\omega(k)) e^{-i\omega(k) t}
+\bar{C}(\omega(k))
e^{i\omega(k) t} \right)\sin\left(\vert k \vert x\right)}{\sqrt{\pi \omega}},
\end{equation}
and in this case the dispersion relation is very simple $\omega=\vert k\vert$
and with this one can easily reconstruct the solution to the original
form of the wave equation before the rescaling.

Let us now consider the discretized version of the Hamiltonian,
\begin{equation}
H(i) = \frac{P_\phi(i)^2}{2 \epsilon} + \frac{\left(\phi(i+1)-\phi(i)\right)^2}{2\epsilon},\label{hamiltonian}
\end{equation}
where the $\epsilon$ in the first term is a remnant of the fact that the
momentum is a density. The resulting discrete wave equation can be
solved in modes,
\begin{equation}
\phi(j) = \sum_{n=-N}^{N}
\frac{1}{\sqrt{2N {\omega(n)}}}
\left(C(\omega(n)) e^{-i\omega(n) t} +\bar{C}(\omega(n))
e^{i\omega(n) t}\right)  {\rm sgn}(n) \sin\left(\frac{j\pi n}{N}\right),
\end{equation}
where all the sums from $-N$ to $N$ exclude zero since there is a
minimum value for the momentum in a box. The frequencies are given by
$\omega(n) = \vert 2 \sin(\pi n/(2 N))/\epsilon\vert$. For
further computations it is useful to define $p\,(n)\equiv \pi n/L$ and
$L=N\epsilon$ and
\begin{equation}
\phi(n,t) \equiv\frac{1}{\sqrt{\omega(n)}}
\left(C(\omega(n)) e^{-i\omega(n) t} +\bar{C}(\omega(n))
e^{i\omega(n) t}\right)  {\rm sgn}(n).
\end{equation}
The momentum is given by,
\begin{equation}
P_\phi(j) = \sum_{n=-N}^{N}
\frac{i}{\sqrt{2N {\omega(n)}}}
\left(-\omega(n) C(\omega(n)) e^{-i\omega(n) t}+\omega(n)\bar{C}(\omega(n)) e^{i\omega(n) t}
\right) \, {\rm sgn}(n)\sin\left(\frac{j\pi n}{N}\right)\epsilon,
\end{equation}
and we define,
\begin{equation}
P_\phi(n,t) =
{\frac{i}{ \sqrt{\omega(n)}}}
\left(-\omega(n) C(\omega(n)) e^{-i\omega(n) t}+\omega(n)\bar{C}(\omega(n)) e^{i\omega(n) t}
\right)  {\rm sgn}(n)\epsilon,
\end{equation}
One can quantize the fields, with discrete commutation relations,
\begin{equation}
\left[\hat{\phi}(i),\hat{P}_\phi(j)\right]=i\delta_{i,j},
\end{equation}
which naturally lead to the introduction of the creation and annihilation
operators,
\begin{equation}
\left[\hat{C}(\omega(n)),\hat{\bar{C}}(\omega(m))\right]=
\frac{1}{2\epsilon}(\delta_{n,m}+\delta_{n,-m}).
\end{equation}
With this one can
compute the free propagators. The Feynman propagator is given by,
\begin{equation}
G^{(0)}(n,t,n',t')= \langle 0 \vert T\left(\phi(n,t),\phi(n',t')\right) \vert 0 \rangle =
D(n,t,t')
\left(\delta_{n,n'}-\delta_{-n,n'}\right),
\end{equation}
where $T$ is the time ordered product and
\begin{equation}
D(n,t,t')=
\left[
\frac{\Theta(t-t') \exp\left(-i \omega(n) (t-t')\right)}{\epsilon \omega(n)}+
\frac{\Theta(t'-t) \exp\left(-i \omega(n) (t'-t)\right)}{\epsilon \omega(n)}
\right],
\end{equation}
or, using the residue theorem,
\begin{equation}\label{35}
D(n,t,t')=\frac{i}{\pi} \int_{-\infty}^\infty \frac{d\omega}{\epsilon}
\frac{1}{\omega^2 -\omega(n)^2+i\sigma} \exp\left(-i\omega (t'-t)\right)
\end{equation}

The previous expressions were in Fourier space. In direct space, one
has,
\begin{equation}
G^{(0)}(j,t,k,t') = \sum_{n=-N}^N \sum_{n'=-N}^N \frac{1}{N}
\sin\left(\frac{j\pi n}{N}\right)
\sin\left(\frac{k\pi n'}{N}\right)
G^{(0)}(n,t,n',t').
\end{equation}

\subsection{Polymerizing the scalar field}
\label{field}
Having computed the free propagator we now turn to study the
polymerized propagator. We start by noticing that the Hamiltonian can
be rewritten (again ignoring boundary terms) as,
\begin{equation}
H=\sum_i H(i)  = \sum_i \frac{P_\phi(i)^2}{2 \epsilon} + \frac{\left(\phi(i+1)-\phi(i)\right)^2}{2\epsilon}=
\sum_i
\frac{P_\phi(i)^2}{2 \epsilon} - \frac{\left(\phi(i+1)+\phi(i-1)-2\phi(i)\right)\phi(i)}{2\epsilon},
\end{equation}
and the rearrangement makes the expression appear more readily
symmetric in $i+1$ and $i-1$. We proceed to polymerize,
\begin{equation}
H=\sum_i\left(
\frac{P_\phi(i)^2}{2 \epsilon} - \frac{\sin\left(\beta\left(\phi(i+1)
+\phi(i-1)-2\phi(i)\right)\right)\sin(\beta \phi(i))}{2\epsilon\beta^2}\right).
\end{equation}
At this point some comments are in order. There are many possible
choices at the time of polymerizing the theory. For instance, we could
have chosen to polymerize $\phi(i+1)+\phi(i-1)-2\phi(i)$ as we did or
we could have polymerized each term in the sum individually. In the
lattice one can also choose to polymerize the momentum $P_\phi(i)$ (in
the continuum this may be more difficult since $P$ is a
density)\footnote{This is the reason our work is not easily compared
  with that of Hossain, Husain and Seahra \cite{HoHuSe}. They
  polymerize the momentum in the continuum. The density nature of the
  momentum leads them to a polymerization parameter that is
  dimensionful, unlike our case.}. If one polymerizes $P_\phi(i)$, when
one takes the continuum limit, since the continuum momentum is
$P_\phi(i)/\epsilon$, one would get for the first term in the
Hamiltonian $\sin^2(\beta \epsilon P_\phi)/\beta^2\epsilon$ and in the
limit $\epsilon\to 0$ one would recover a non-polymerized theory and
therefore we would not be making contact with usual loop quantum
gravity results.  Polymerizing the fields as we have chosen yields in the
continuum limit a term $\phi''(x)\sin(\beta \phi(x))/\beta$ showing
that the continuum theory is polymerized. It is interesting to notice
that spatial derivatives of fields are well defined in the Bohr
compactification even if the field operators themselves are not. In this
section we will work with a polymerization of the field rather than of
the momentum.  In a discrete theory polymerizing either fields or
momenta is possible, but it does not lead to equivalent theories. For
completeness, in the next section we will discuss the theory that
results from polymerizing the momenta. In previous treatments in the
continuum \cite{thiemannscalar} the scalar field has been polymerized,
although in the case of the harmonic oscillator, which one can
consider closely related to a scalar field, a polymerization of the
momentum has been preferred \cite{harmonicoscillator}.

We are going to work perturbatively, expanding in $\beta$. The Hamiltonian
we will consider is $H=H_0+H_{\rm int}$ with
\begin{equation}
H_0=\sum_i\left(
\frac{P_\phi(i)^2}{2 \epsilon} -
\frac{\phi(i)\left(\phi(i+1)+\phi(i-1)-2\phi(i)\right)}{2\epsilon}\right),
\end{equation}
and
\begin{equation}
H_{\rm int} = \sum_i
\frac{1}{2\epsilon}\left(
\frac{1}{6}\phi(i) \left(\phi(i+1)+\phi(i-1)-2\phi(i)\right)^3
+\frac{1}{6}\phi(i)^3
\left(\phi(i+1)+\phi(i-1)-2\phi(i)\right)\right)\beta^2.
\end{equation}
This interaction Hamiltonian comes from expansion in beta and keeping
the first two leading terms. With it we compute the interacting
propagator to leading order,
\begin{equation}
G^{(2)}(j,t,k,t')= G^{(0)}(j,t,k,t')+ \frac{i^2}{2!} \int_{-\infty}^\infty dt_1
\int_{-\infty}^\infty dt_2
\sum_{j'=-N}^N \sum_{k'=-N}^N
\langle 0 \vert T\left(\phi(j,t)\phi(k,t') H_{\rm int}(j',t_1) H_{\rm int}(k',t_2) \right)\vert 0 \rangle
\end{equation}
To compute this expression
it is convenient to rewrite the interaction Hamiltonian in momentum
space (we use letters up to $k$ for the field representation and
letters starting with $m$ for the momentum representation)
\begin{eqnarray}
H_{\rm int}(j',t_1) &=&
\sum_{n,m,p,q=-N}^N
\left\{
\frac{1}{48 N^2} \beta^2 \epsilon^5
\sin\left(\frac{\pi j' n}{N}\right) \phi(n,t_1)
\omega(m)^2
\sin\left(\frac{\pi j' m}{N}\right)
\phi(m,t_1)\right.\\
&&\times
\omega(p)^2
\sin\left(\frac{\pi j' p}{N}\right)
\phi(p,t_1)
\omega(q)^2
\sin\left(\frac{\pi j' q}{N}\right)
\phi(q,t_1)\nonumber\\
&&+\frac{1}{48 N^2}\beta^2 \epsilon
\sin\left(\frac{\pi j' n}{N}\right)
\phi(n,t_1)
\sin\left(\frac{\pi j' m}{N}\right)
\phi(m,t_1)\nonumber\\
&&\times\left.
\sin\left(\frac{\pi j' p}{N}\right)
\phi(p,t_1)
\omega(q)^2
\sin\left(\frac{\pi j' q}{N}\right)
\phi(q,t)\right\}.\nonumber
\end{eqnarray}
We now use the identity,
\begin{eqnarray}
\Delta(n,m,p,q) &\equiv&
\sum_{j'=-N}^N \frac{4}{N^2}
\sin\left(\frac{\pi j' n}{N}\right)
\sin\left(\frac{\pi j' m}{N}\right)
\sin\left(\frac{\pi j' p}{N}\right)
\sin\left(\frac{\pi j' q}{N}\right)\\
&=&
\frac{1}{N} \left[
 \delta_{n+m,p+q}
+\delta_{n+p,m+q}
+\delta_{n+q,m+p}
+\delta_{n+m+p+q}
-\delta_{n,m+p+q}
-\delta_{m,n+p+q}
-\delta_{p,n+m+q}
-\delta_{q,n+m+p}\right].\nonumber
\end{eqnarray}
We can use this identity to get,
\begin{equation}
\sum_{j'=-N}^N H_{\rm int}(j',t_1) =
\frac{1}{192} \sum_{n,m,p,q=-N}^N
\phi(n,t_1) \phi(m,t_1)
\phi(p,t_1) \phi(q,t_1)
\left[\left(\omega(m)^2\omega(p)^2\epsilon^4+1\right)\epsilon\,
 \omega(q)^2\right] \beta^2 \Delta(n,m,p,q),
\end{equation}
where we will use the notation,
\begin{equation}
f(m,p,q) =\left[\left(\omega(m)^2\omega(p)^2\epsilon^4+1\right)\epsilon\,
 \omega(q)^2\right].
\end{equation}
Putting everything together we get,
\begin{eqnarray}
G^{(2)}(n_1,t_1,n_2,t_2) &=&
G^{(0)}(n_1,t_1,n_2,t_2)
+\frac{i^2}{2!} \langle 0 \vert T\big(\phi(n_1,t_1)\phi(n_2,t_2)\\
&&\times
\frac{1}{192}
\int_{-\infty}^\infty dt'
\sum_{n,m,p,q=-N}^N
:\phi(n,t') \phi(m,t') \phi(p,t')\phi(q,t') :
f(n,m,p)\beta^2\Delta(n,m,p,q)\nonumber\\
&&\times \frac{1}{192}
\int_{-\infty}^\infty dt''
\sum_{n',m',p',q'=-N}^N
:\phi(n',t'') \phi(m',t'') \phi(p',t'')\phi(q',t'') :
f(n',m',p')\beta^2\Delta(n',m',p',q') \big)\vert 0\rangle.\nonumber
\end{eqnarray}
Using Wick's theorem the above expression can be rewritten as
a sum of diagrams of the form,
\begin{eqnarray}
G^{(2)}(n_1,t_1,n_2,t_2) &=&
G^{(0)}(n_1,t_1,n_2,t_2)
-\frac{32}{3N^2} \sum_{m,p=-N}^N\int_{-\infty}^\infty dt'dt''
\left[D(n_1,t_1,t')
D(m,t',t'')\right.\\
&&\times\left. D(p,t',t'')
D(n+m-p,t',t'')
D(n_2,t'',t_2)\right]f^2(m,p,n+m-p)\beta^4
\left(\delta_{n_1,n_2}-\delta_{n_1,-n_2}\right),\nonumber
\end{eqnarray}
or, graphically,

\includegraphics[height=3cm]{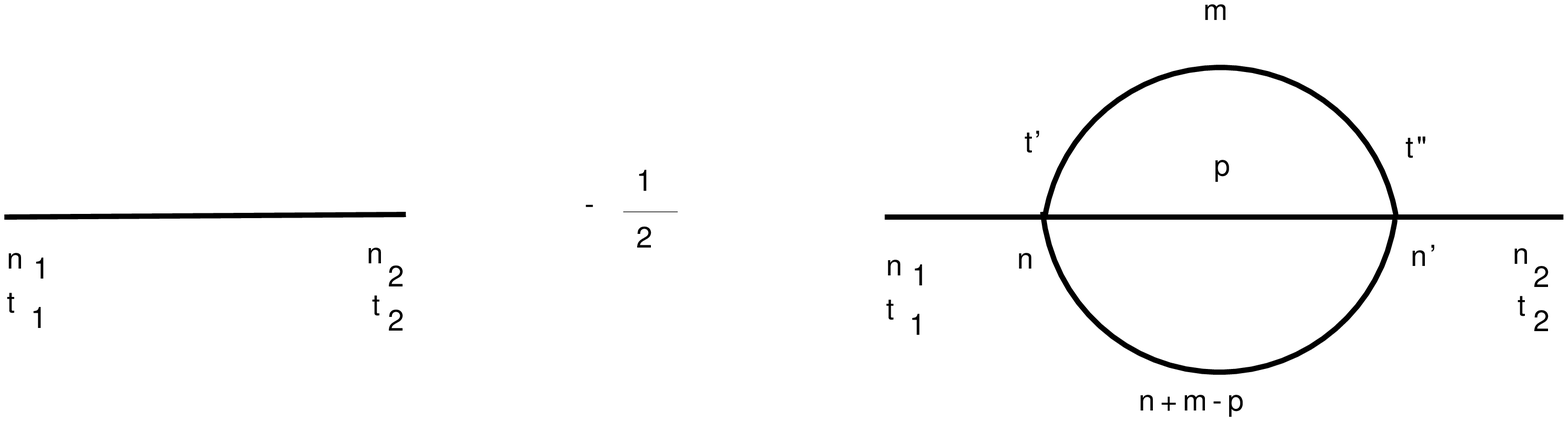}

It is now convenient to Fourier transform in time,
\begin{eqnarray}
G^{(2)}(n_1,\omega_1,n_2,\omega_2) &=&
\frac{4 \pi i}{\epsilon} \frac{1}{\omega_1^2-\omega(n_1)^2+i\sigma}
\delta(\omega_1-\omega_2)\left(\delta_{n_1,n_2}-\delta_{n_1,-n_2}\right)\\
&&-\frac{32}{3}
\frac{1}{2}
\frac{4 \pi i}{\epsilon\left(\omega_1^2-\omega(n_1)^2+i\sigma\right)}
\sum_{m,p=-N}^N  \int_{-\infty}^\infty d\omega' d\omega'' \nonumber\\
&&\times
\frac{4 \pi i}{\epsilon\left((\omega')^2-\omega(m)^2+i\sigma\right)}
\frac{4 \pi i}{\epsilon\left((\omega'')^2-\omega(p)^2+i\sigma\right)}\nonumber\\
&&\times
\frac{4 \pi i}{\left(\left(\omega_1-\omega'-\omega''\right)^2
-\omega({n_1-m-p})^2+i\sigma\right)} f^2(m,p,n_1-m-p)\beta^4
\nonumber\\
&&\times
\frac{4 \pi i}{\left(\omega_2^2-\omega(n_2)^2+i\sigma\right)}
\delta(\omega_1-\omega_2)
\left(\delta_{n_1,n_2}-\delta_{n_1,-n_2}\right),\nonumber
\end{eqnarray}
and the sums can be converted to integrals. Care should be exercised
not to allow the denominators to vanish, since in the original
discrete expression the denominators did not vanish. We recall that
$\omega(n) = \vert 2 \sin(\pi n/(2 N))/\epsilon\vert$ and $p(n)\equiv
\pi n/L$ with $L=N\epsilon$. Therefore $\omega(n)=2\vert \sin(\epsilon p(n)/2)
\vert/\epsilon \sim p(n)$. One then approximates,
\begin{equation}
\sum_{m=1}^N \to \frac{L}{\pi} \int_{\pi/L}^{\pi/\epsilon} dp
\end{equation}
and the sum from $-N$ to 1 takes an analogous form. The expression for the
Green function up to second order is,
\begin{eqnarray}
G^{(2)}(n_1,\omega_1,n_2,\omega_2) &=&
\frac{4 \pi i}{\epsilon} \frac{1}{\omega_1^2-p(n_1)^2+i\sigma}
\delta(\omega_1-\omega_2)\left(\delta_{n_1,n_2}-\delta_{n_1,-n_2}\right)\\
&&-\frac{32}{3}\frac{1}{2}
\frac{4 \pi i}{\epsilon\left(\omega_1^2-p(n_1)^2+i\sigma\right)}
\frac{1}{\pi^2}\int_{-\infty}^\infty d\omega' d\omega''
\left[\int_{-\pi/\epsilon}^{-\pi/L}+\int_{\pi/L}^{\pi/\epsilon}\right]
dp_1 dp_2\nonumber\\
&&\times
\frac{4 \pi i}{\epsilon\left((\omega')^2-p_1{}^2+i\sigma\right)}
\frac{4 \pi i}{\epsilon\left((\omega'')^2-p_2{}^2+i\sigma\right)}\nonumber\\
&&\times
\frac{4 \pi i}{\left(\left(\omega_1-\omega'-\omega''\right)^2
-p({n_1-p_1-p_2})^2+i\sigma\right)} \tilde{f}^2(p_1,p_2,p(n_1)-p_1-p_2)\beta^4\nonumber\\
&&\times
\frac{4 \pi i}{\left(\omega_2^2-\omega(n_2)^2+i\sigma\right)}
\delta(\omega_1-\omega_2) \left(\delta_{n_1,n_2}
-\delta_{n_1,-n_2}\right)\nonumber
\end{eqnarray}
where
\begin{equation}
\tilde{f}(p_1,p_2,p(n_1)+p_1-p_2) =
\left(\epsilon^4\left(p_1{}^2p_2{}^2\right)+1\right)\left(p(n_1)
+p_1-p_2\right)^2\epsilon^2
\end{equation}
The integrals can be computed by an analytic extension to the
Euclidean theory and by carrying out an expansion in $p \epsilon$, the
next expression is correct up to order $O(\epsilon^4 p^4)$. That is,
we are assuming the wavelength of the scalar field is much larger than
the lattice spacing. If one takes into account powers higher than $p
\epsilon$ one has higher corrections in powers of $p$. The result, not
including those terms, is,
\begin{eqnarray}
G^{(2)}(n_1,\omega_1,n_2,\omega_2) &=&
G^{(0)}(n_1,\omega_1,n_2,\omega_2) +
\left[\frac{\alpha_1 \beta^4}{\epsilon^2} +
\beta^4 \alpha_2 p(n_1)^2\right]
\frac{4\pi i}{\epsilon} \frac{\delta(\omega_1-\omega_2)
\left(\delta_{n_1,n_2}-\delta_{n_1,-n_2}\right)}
{\left(\omega_1^2-p(n_1)^2+i\sigma\right)^2}\label{55}\\
&=&
\frac{4 \pi i}{\epsilon}
\frac{1}{\omega_1^2-p(n_1)^2\left(1+\alpha_2 \beta^4\right)-
\frac{\alpha_1\beta^4}{\epsilon^2} + i\sigma}
\left(\delta_{n_1,n_2}-\delta_{n_1,-n_2}\right)\delta(\omega_1-\omega_2)
\end{eqnarray}
where $\alpha_{1}$ and $\alpha_2$ are constants of order one, and the
last expression yields (\ref{55}) if one expands assuming $\beta^4$ is
small.

\subsection{Polymerizing the momentum of the field}

\label{momentum}
We now discuss the choice of polymerizing the momentum.
As before, we write the Hamiltonian as,
\begin{equation}
H=
\sum_i
\frac{P_\phi(i)^2}{2 \epsilon} - \frac{\left(\phi(i+1)+\phi(i-1)-2\phi(i)\right)\phi(i)}{2\epsilon},
\end{equation}
We proceed to polymerize,
\begin{equation}
H=\sum_i
\frac{\sin^2\left(\beta P_\phi(i)\right)}{2 \beta^2 \epsilon} - \frac{\left(\phi(i+1)+\phi(i-1)-2\phi(i)\right)\phi(i)}{2\epsilon},
\end{equation}

As before, we work perturbatively, expanding in $\beta$. The Hamiltonian
we will consider is $H=H_0+H_{\rm int}$ with
\begin{equation}
H_{\rm int}(i) = -\frac{1}{6\epsilon } \beta^2 P_\phi(i)^4.
\end{equation}

And we can now write the Green function up to second order,
\begin{equation}
G^{(2)}(j,t,k,t')= G^{(0)}(j,t,k,t')+ \frac{i^2}{2!} \int_{-\infty}^\infty dt_1
\int_{-\infty}^\infty dt_2
\sum_{j'=-N}^N \sum_{k'=-N}^N
\langle 0 \vert T\left(\phi(j,t)\phi(k,t') H_{\rm int}(j',t_1) H_{\rm int}(k',t_2) \right)
\vert 0 \rangle
\end{equation}
We now rewrite,
\begin{equation}
\sum_{j'=-N}^N H_{\rm int}(j',t_1) =
-\frac{1}{96\epsilon}\sum_{n,m,p,q=-N}^N
P_\phi(n,t')P_\phi(m,t')P_\phi(p,t')P_\phi(q,t')\Delta(n,m,p,q) \beta^2
\end{equation}
Putting everything together we get,
\begin{eqnarray}
G^{(2)}(n_1,t_1,n_2,t_2) &=&
G^{(0)}(n_1,t_1,n_2,t_2)
+\frac{i^2}{2!} \langle 0 \vert T\big(\phi(n_1,t_1)\phi(n_2,t_2)\\
&&\times
\frac{1}{96}
\int_{-\infty}^\infty \frac{dt'}{\epsilon}
\sum_{n,m,p,q=-N}^N
:P_\phi(n,t') P_\phi(m,t') P_\phi(p,t')P_\phi(q,t') :
\beta^2\Delta(n,m,p,q)\nonumber\\
&&\times \frac{1}{96}
\int_{-\infty}^\infty \frac{dt''} {\epsilon}
\sum_{n',m',p',q'=-N}^N
:P_\phi(n',t'') P_\phi(m',t'') P_\phi(p',t'')P_\phi(q',t'') :
\beta^2\Delta(n',m',p',q')\big) \vert 0\rangle.\nonumber
\end{eqnarray}

If we now use Wick's theorem as we did before, there will appear
contractions not only of $\phi$ with itself, but also between $\phi$
and its momentum. Taking into account that the momentum is
related to the derivative of the field $P_\phi=\epsilon \dot{\phi}$
one can compute the expectation values of products of the field
and momentum or products of the momenta by taking derivatives
of (\ref{35}) with respect to time.
\begin{eqnarray}
G^{(2)}(n_1,t_1,n_2,t_2) &=&
G^{(0)}(n_1,t_1,n_2,t_2)
-128 \frac{\beta^4 }{3N^2\epsilon^2}\sum_{m,p,q,m',p',q'=-N}^N
\int_{-\infty}^\infty dt' dt''
D_{\phi P_\phi}(n_1,t_1,m,t')
\\
&&\times
D_{P_\phi P_\phi}(p,t',p',t'')
D_{P_\phi P_\phi}(q,t',q',t'')
D_{P_\phi P_\phi}(m+p-q,t',m'+p'-q',t'')
D_{P_\phi \phi}(m',t'',n_2,t_2)\nonumber
\end{eqnarray}
where
\begin{eqnarray}
D_{\phi\phi}(n_1,t_1,n_2,t_2) &=&
\frac{iL^2}{\pi \epsilon} \int_{-\infty}^\infty \frac{d\omega}{\omega^2-\omega(n_1)^2+i \sigma} \exp\left(-i\omega(t_2-t_1)\right)
\left(\delta_{n_1,n_2}-\delta_{n_1,-n_2}\right)\\
D_{P_\phi\phi}(n_1,t_1,n_2,t_2) &=&
-\frac{L^2}{\pi} \int_{-\infty}^\infty \frac{\omega(n_1)\, d\omega}{\omega^2-\omega(n)^2+i \sigma} \exp\left(-i\omega(t_2-t_1)\right)
\left(\delta_{n_1,n_2}-\delta_{n_1,-n_2}\right)\\
D_{\phi P_\phi}(n_1,t_1,n_2,t_2) &=&
\frac{L^2}{\pi} \int_{-\infty}^\infty \frac{\omega(n_1)\, d\omega}{\omega^2-\omega(n)^2+i \sigma} \exp\left(-i\omega(t_2-t_1)\right)
\left(\delta_{n_1,n_2}-\delta_{n_1,-n_2}\right)\\
D_{P_\phi P_\phi}(n_1,t_1,n_2,t_2) &=&
-\frac{iL^2\epsilon}{\pi } \int_{-\infty}^\infty \frac{\omega(n_1)^2\,d\omega}{\omega^2-\omega(n_1)^2+i \sigma} \exp\left(-i\omega(t_2-t_1)\right)
\left(\delta_{n_1,n_2}-\delta_{n_1,-n_2}\right)
\end{eqnarray}
or, graphically,

\includegraphics[height=3cm]{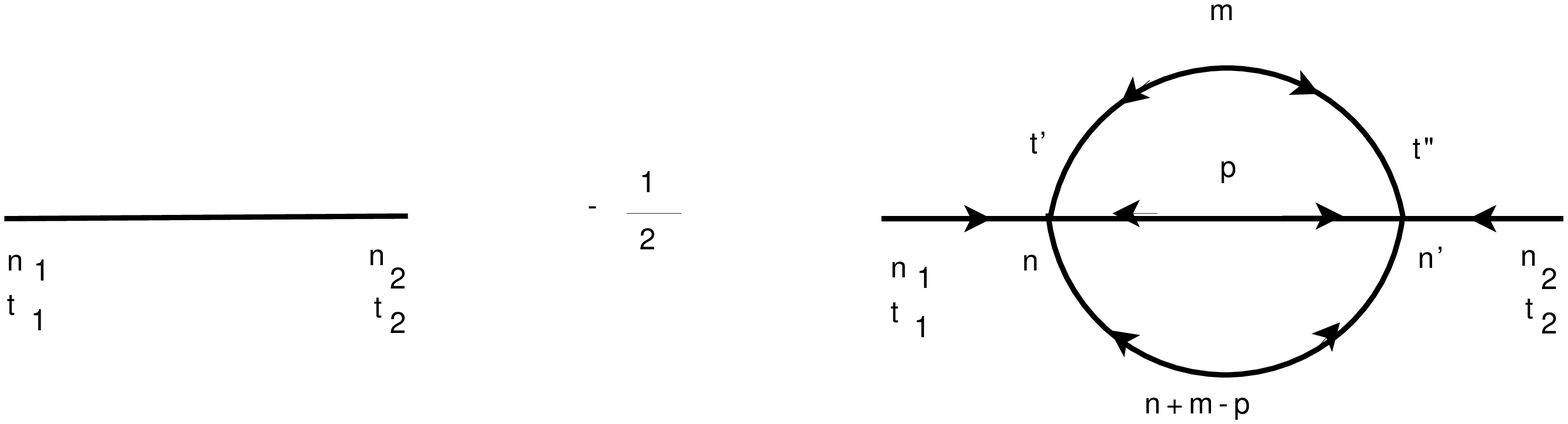}

where the direction of the arrows depend on the order of
appearance of $\phi$ and $P_\phi$ in their product,
meaning an arrow to the right is $\phi P_\phi$, an arrow to the left
$P_\phi \phi$, two arrows mean $P_\phi P_\phi$, and no arrow
means $\phi \phi$.

We Fourier transform in time and take the continuum approximation
for the sums in $p$ and $q$,
\begin{eqnarray}
G^{(2)}(n_1,\omega_1,n_2,\omega_2) &=&
\frac{4 \pi i}{\epsilon} \frac{1}{\omega_1^2-p(n_1)^2+i\sigma}
\delta(\omega_1-\omega_2)\left(\delta_{n_1,n_2}-\delta_{n_1,-n_2}\right)+\\
&&\frac{128}{3}
\frac{1}{2\pi^2}
\frac{1}{\left(\omega_1^2-p(n_1)^2+i\sigma\right)}
\frac{1}{\pi^2}\int_{-\infty}^\infty d\omega' d\omega''
\left[\int_{-\pi/\epsilon}^{-\pi/L}+\int_{\pi/L}^{\pi/\epsilon}\right] dp_1 dp_2
\left(\frac{i}{2\pi^3}\right)^3
\nonumber\\
&&\times
\frac{\epsilon p_1^2}{\left((\omega')^2-p_1{}^2+i\sigma\right)}
\frac{\epsilon p_2^2}
{\left((\omega'')^2-p_2{}^2+i\sigma\right)}\nonumber\\
&&\times
\frac{\left(p(n_1)-p_1-p_2\right)^2}
{\left(\left(\omega_1-\omega'-\omega''\right)^2
-(p(n_1)-p_1-p_2)^2+i\sigma\right)}
\beta^4\nonumber\\
&&\times
\frac{4 \pi i \omega(n_1)^2}{\left(\omega_2^2-\omega(n_2)^2+i\sigma\right)}
\delta(\omega_1-\omega_2) \left(\delta_{n_1,n_2}
-\delta_{n_1,-n_2}\right)\nonumber
\end{eqnarray}
The integrals can be computed as before, expanding
in $p\, \epsilon$ and analytically continuing to the Euclidean theory,
\begin{eqnarray}
G^{(2)}(n_1,\omega_1,n_2,\omega_2) &=&
G^{(0)}(n_1,\omega_1,n_2,\omega_2)
+
\beta^4 \alpha_2 p(n_1)^2
\frac{4\pi i}{\epsilon} \frac{\delta(\omega_1-\omega_2)
\left(\delta_{n_1,n_2}-\delta_{n_1,-n_2}\right)}
{\left(\omega_1^2-p(n_1)^2+i\sigma\right)^2}\label{73}\\
&=&
\frac{4 \pi i}{\epsilon}
\frac{1}{\omega_1^2-p(n_1)^2\left(1+\alpha_2 \beta^4\right)
 + i\sigma}\left(\delta_{n_1,n_2}-\delta_{n_1,-n_2}\right)\delta(\omega_1-\omega_2)\label{92}
\end{eqnarray}
where the last expression yields (\ref{73}) if one expands assuming
$\beta^4$ is small.

\section{Lorentz invariance}

The derived propagators violate Lorentz invariance. It is therefore
worthwhile discussing in some detail the nature of the
violation. There are two distinct origins for it, which we will
discuss in the following two subsections.

\subsection{Choices of polymerization}

First of all, one has violation of Lorentz invariance due to the
polymerization. This can be seen in terms like the dispersion relation
implied by the denominator of
(\ref{92}),
\begin{equation}
\omega_1^2-p(n_1)^2\left(1+\alpha_2 \beta^4\right).
\end{equation}
It should be noted that these terms depend on the value of the
polymerization parameter $\beta$. The order in $\beta$ at which these
terms appear depends on choices made at the time of polymerization. To
see this, let us write the polymerized momentum term in Hamiltonian as
\begin{equation}
\frac{c\sin(\beta
  P_\phi(i))}{\sqrt{2\epsilon}\beta}+\frac{(1-c)\sin(3\beta P_\phi(i)}
{3\sqrt{2\epsilon}\beta}\label{high-beta-1}
\end{equation}
and try to find $c$ such that we are left only with the
non-perturbative term in $P_\phi(i)$ and a perturbative term in
$\beta^4$, thus neglecting the $\beta^2$ term. This way we can analyze
just the effects of $\beta^4$ order term in the propagator. Expanding
(\ref{high-beta-1}) in $\beta$ we get
\begin{equation}
  \frac{P_\phi(i)^2}{2\epsilon}+\left(\frac{4}{3}\frac{c
      P_\phi(i)^4}{\epsilon}
-\frac{3}{2} \frac{P_\phi(i)^4}{\epsilon}\right) \beta^2
+\left(\frac{8}{9}\frac{c^2 P_\phi(i)^6}{\epsilon}
-\frac{8}{3}\frac{c P_\phi(i)^6}{\epsilon}
+ \frac{9}{5} \frac{P_\phi(i)^6}{\epsilon}\right) \beta^4.
\end{equation}
Obviously from the coefficient of $\beta^2$ we see that setting
$c=\frac{9}{8}$, the $\beta^2$ order term cancel and we are left with
only the non-perturbative term and a perturbative term in $\beta^4$
which is
\begin{equation}
H_{\rm int}(i)=-\frac{3}{40\epsilon}P_\phi(i)^6 \beta^4.
\end{equation}
This would lead to corrections of order $\beta^8$ instead of $\beta^4$
in (\ref{92}).  We therefore see that the order in $\beta$ at which
corrections appear can be shifted arbitrarily by choosing suitable
polymerizations of the theory and therefore, assuming that the
polymerization parameter is small, one can make the corrections as
small as desired.

It is interesting to emphasize that these corrections do not
necessarily involve the Planck length. Unlike when one polymerizes the
gravitational variables, there is no a priori reason to relate the
polymerization parameter of a scalar field to the quantum of area. The
reason for having a relation between the parameter and the area
in the gravitational variables is
because one is dealing with a true holonomy along a spatial loop which
encloses an amount of area. In the case of the scalar field however,
one is dealing with point holonomies and therefore they do not enclose
area.
These polymerization dependent Lorentz
violations are also not of the form conjectured by Ho\v{r}ava in his
``gravity at the Lifshitz point'', \cite{horava} since there the
corrections were Planck scale dependent. These corrections amount to a
redefinition of the speed of light for a massless scalar field. This
could lead to experimental problems if similar redefinitions do not
occur for other fields, since differences in the speed of propagation
of massless fields is severely constrained experimentally.

In addition to the Lorentz violations due to the polymerization there
is the issue that we are working in a discrete theory for which we have
failed to find a continuum limit. Since we have been unable to find a
ground state for which the master constraint is zero, we worked with a
variationally found state that minimized the master constraint. The
minimum found was achieved with a finite lattice spacing. We found
that the minimum of the master constraint occurs at a spacing large
compared to the Planck length but small compared to particle physics
scalings. This finiteness of the lattice implies that expressions like
(\ref{92}) are only approximate and there are corrections that go as
the momentum to the fourth power times the lattice spacing
squared. Those corrections are of Ho\v{r}ava type. This is good since
Ho\v{r}ava has argued that such corrections help make theories
finite, as one would expect in a lattice treatment like the
one we pursue.

A last point here is that we have discussed the corrections due to
polymerization and  due to the lattice discreteness separately
where in reality they are not separate. When we studied the corrections
due to polymerization we took the continuum limit to get simple
expressions. In reality, if one kept on working on the lattice till
the end there would appear terms involving the lattice spacing as
well in the corrections due to polymerization.

\subsection{The arguments of Collins et al.}

Collins et al. \cite{collins} have argued that Lorentz violations of
the second kind considered in the previous section (more precisely, the
corrections that depend on the Planck scale) could lead to
unacceptably large effects when one considers interactions at one loop
level. One has to be careful in applying their arguments directly to
the propagators we discussed in the previous section since they were
derived in the low energy limit, expanding in powers of the momentum.
It is of course not legitimate to expand something in powers of a
given parameter, keeping the lowest order, and then evaluating the
expression for large values of the expansion parameter. So in order to
reconstruct the argument of Collins et al. for our case, we would need
the full expression of the propagators, which we do not have. One can
sketch how it is likely to go. The terms we have neglected are due to
the fact that the theory is put on a lattice. One knows that the
lattice propagator for a scalar field takes the form,
\begin{equation}
  \frac{1}{-m^2+\omega^2-\sum_i \frac{\sin^2\left(a p_i\right)}{a^2}},
\end{equation}
where $a$ is the lattice spacing.  The presence of the sine function
implies that for large values of the momentum, the quantity remains
finite. In particular, if one re-does the calculations of Collins et
al. with such propagators, one finds that it leads to large Lorentz
violations, due the asymmetric treatment of space and time.  However,
it would be hasty to conclude that this is a problem. The reason for
this is that in general relativity one is not supposed to take the
zeroth component of the coordinates as a time variable. What we have
done in this paper is to work out a gauge fixed quantization of the
scalar field, which implies a specific choice of coordinates
classically. However, in a generally covariant theory one should
really use physical systems of reference through the introduction of
real ``clocks and rods''. This, in turn solves the ``problem of time''
of such theories. In a more realistic calculation than the one done
here one would have other matter fields present that can be used as
``clocks and rods'' to measure time and space.

In generally covariant theories one should construct and ask physical
questions about observables. This, in particular, applies to the
calculation of Green's functions, as was already noted by DeWitt
\cite{dewitt}. These would not correspont to the propagators we have
considered up to now here, which are constructed in terms of the
coordinates, but would have to be cast in terms of the times and
distances measured by the physical ``clocks and rods'', so the
resulting propagators are Dirac observables,
\begin{equation}
  D\left(T_1,T_1,\vec{X}_1\vec{X}_2\right)=\int dt_1 \int d^3x_1 \int dt'{}_1 \int
  d^3x'{}_2 D\left(t_1,t_2,\vec{x}_1\vec{x}_2\right) 
{\cal P}\left(t_1,T_1\right)
{\cal P}\left(t_2,T_2\right)
{\cal P}\left(\vec{x}_1,\vec{X}_1\right)
{\cal P}\left(\vec{x}_2,\vec{X}_2\right),
\end{equation}
and we are considering a situation where space is locally flat,
otherwise the integrals should involve square roots of the determinant
of the metric. The above expression can be derived in detail in the
discussion of the problem of time starting from conditional
probabilities of evolving Dirac observables, see \cite{time,obregon}
for details. The quantities ${\cal P}$ are probability distributions
that tell us what is the chance that given a value of a variable, say,
$t_1$ the real clock is measuring $T_1$ and similarly for the
others. Generically, these probability distributions will be highly
peaked ---provided one chose sensible clocks and rods---, indicating
that there is little difference between the parameter time $t_1$ and
the clock time $T_1$ and similarly for the spatial rods. The width of
the distributions will depend on the physical details of the clocks
and rods chosen, but the important point is that there exist
fundamental physical limitations that dictate that the widths cannot
be arbitrarily small \cite{ng}. These limitations in fact state that the widths
should be considerably larger than the Planck scale (and of the
lattice spacing considered in this paper). This introduces naturally a
cutoff in four-momentum space that implies that the Lorentz violating
contributions we encountered above will be suppressed.

For instance, let us just concentrate on the effect of the clock, as
the one from the rods is similar. We assume that ${\cal P}(t,T)$ is an
approximation of the Dirac delta given by a step function of width
$2\sigma$. In that case one can carry out the integral explicitly to
get,
\begin{equation}
D\left(T,\vec{x},T',\vec{x}'\right)    =
\int_{-\pi/a}^{\pi/a} d^3p \frac{e^{i\vec{p}\cdot \vec{x}}}
{2\omega_a} \frac{\sin^2\left(\omega_a \sigma\right)}{\omega_a^2
  \sigma^2} e^{-i\omega_a \vert T-T'\vert},
\end{equation}
where $\omega_a=\sqrt{m^2 +\sum_j\frac{\sin^2\left(a
      p_j\right)}{a^2}}$.  In usual quantum gravity scenarios,
$\sigma$ is proportional to some power of the Planck length and grows
with time \cite{obregon}. We notice that this introduces an
ultraviolet cutoff in $\omega_a$.

It is worthwhile discussing how does Lorentz invariance emerge in a
context like this (a point emphasized by Rovelli and Speziale in
\cite{rovelli}). When one introduces a set of clocks and rods
$(T,\vec{X})$, one is manifestly breaking Lorentz invariance. The
latter is recovered in the sense that if one takes the same set of
clocks and rods as before (or one similarly constructed) and boosts it
$(T',\vec{X}')$ and one carries out the above calculation one will get
that the physical propagator is invariant,
\begin{equation}
  D\left(T_1,T_2,\vec{X}_1,\vec{X}_2\right)=
  D\left(T_1',T_2',\vec{X}_1',\vec{X}_2'\right).
\end{equation}
In order for this equality to hold we need two conditions: that the
uncertainties in the clocks and the rods be the same (otherwise that
would automatically violate Lorentz invariance), and the second one that
for small values of the momentum the propagators considered have the
usual Lorentz invariant form.

\section{Discussion}

We have studied the propagator in a polymerized scalar field theory
with spherical symmetry.  This requires defining a vacuum, which we
took to be the Fock vacuum based on our experiences in Paper I. In
this paper we further confirmed that this vacuum is adequate by
computing the expectation value of the master constraint polymerized and
expanded to leading order in the polymerization parameter and noting
that the corrections introduced in the expectation value by the
polymerization are very small. We then proceeded to study the
propagator to leading order in the polymerization parameter for two
different choices of polymerization: either polymerizing the field or
its canonically conjugate momentum. We ended up with propagators that
had Lorentz violations of two different types, one stemming from the
polymerization of the scalar field and the other from the
discreteness that is remnant from the uniform discretization procedure,
since the state that minimizes the expectation value of the master
constraint does so for a finite lattice spacing. This could be a
temporary limitation until a better state is found, or it could well be
that such a state actually does not exist.

The Lorentz violation due to polymerization can be made arbitrarily
small by a suitable choice of the polymerization parameter. This is
because in the case of a scalar field this parameter is not obviously
associated with an area and therefore not limited by the minimum area
eigenvalue as is the case for gravitational variables. The order of
the violation in the parameter can also be changed by choices in
polymerization.  We also argued that the Lorentz violations that arise
due to the use of a lattice are not of the type considered by Collins
et al. and that if one uses real clocks and rods to characterize
space-time points in such a way that propagators are Dirac
observables, potential divergences in the integrals on the frequencies
are contained and may not lead to large Lorentz violations either.

\section*{Acknowledgments}

We wish to thank Abhay Ashtekar, John Collins, Alejandro P\'erez, Joe
Polchinski, Rafael Porto, Carlo Rovelli and Daniel Sudarsky for
discussions.  This work was supported in part by grant
NSF-PHY-0650715, funds of the Hearne Institute for Theoretical
Physics, FQXi, CCT-LSU, Pedeciba and ANII PDT63/076. This publication
was made possible through the support of a grant from the John
Templeton Foundation. The opinions expressed in this publication are
those of the author(s) and do not necessarily reflect the views of the
John Templeton Foundation.

\section*{Appendix}

Since the construction carried out is rather elaborate, it is
worthwhile spelling out for the readers the various assumptions made
to reach the results of this paper.

a) We started in Paper I by attempting to find a ``vacuum'' for the
combined gravity and scalar field system in spherical symmetry on a
lattice. We polymerized the gravitational variables but as a first
exploration we kept the scalar field not polymerized. By vacuum in
this context we mean a state that minimizes the master constraint on
the lattice with eigenvalue zero but at the same time that minimizes
the energy of the matter Hamiltonian. We chose the simplest factor
ordering of the master constraint which made it self-adjoint.

b) To find the vacuum we proceeded variationally, using a trial
state. The trial state consisted of Gaussians at each lattice point
centered around the classical solution (flat space with a deficit
angle) for the gravitational variables. For the scalar field the trial
state was a Fock state modified by the gravitational background and
quantum corrections. The deficit angle arises due to the zero point
energy of the scalar field, which in $1+1$ dimensions does not yield a
local curvature (a cosmological constant as in 4 dimensions) and only
produces global curvature.  The total trial state was assumed to be
the direct product of the Gaussians and the Fock state. For
calculations involving the vacuum it is reasonable to assume no
correlations. This assumption  would be wrong for excited states.

c) We proceeded to minimize the master constraint given the
variational trial state. The minimization parameters were the widths
of the Gaussians for the gravitational variables. We found that the
minimum was achieved with essentially constant values for the
parameters along the lattice. The minimum value of the master
constraint did not correspond to the lattice spacing going to zero.
In fact the expectation value diverges in that limit. The theory does
not have a continuum limit and the minimum of the master constraint is
not zero, but is very small and so it is the ideal lattice spacing. It is
large compared to the Planck length but still very small for particle
physics standards. This concluded Paper I.

d) In this paper we wished to study the propagator of the polymerized
scalar field. First we had to convince ourselves that the vacuum of
Paper I was still useful, since it was derived without the assumption that
the field was polymerized. We re-evaluated the expectation value of
the master constraint with the scalar field polymerized on the vacuum of
Paper I. We  found that the value differed very little from the one
found in Paper I. This validated the use of the vacuum of Paper I in the
polymerized context.

e) We studied the polymerized scalar field treating the polymerization
parameter as being small, thinking of the un-polymerized theory as a
``free'' theory and the extra terms stemming from the polymerization
as perturbations. We then applied standard quantum field theory techniques
to find the propagator.

f) However, we were on the lattice, so we used certain approximations
to evaluate summations by taking the continuum limit and evaluating
integrals. This way, to leading order in the lattice spacing the
approximation is good.

g) We found that the propagators acquired Lorentz violating
corrections due to the polymerization. Some of the terms not
explicitly evaluated in the approximation f) also lead to Lorentz
violations.

h) We discussed if the Lorentz violations could cause problems as
those discussed by Collins et al. We argued that it may be possible
that they do not, if one carries out a proper treatment of the problem
of time and loop quantum gravity ends up providing a regularization of
the matter fields with certain properties.


\begin{thebibliography}{99}

\bibitem{nosotros}
  R.~Gambini, J.~Pullin and S.~Rastgoo,
  Class.\ Quant.\ Grav.\  {\bf 26}, 215011 (2009)
  [arXiv:0906.1774 [gr-qc]].

\bibitem{husain}
W. Unruh, Phys. Rev. D14, 870 (1976); V. Husain, O. Winkler,
  Phys.\ Rev.\  D {\bf 71}, 104001 (2005)
  [arXiv:gr-qc/0503031].

\bibitem{uniform}
  M.~Campiglia, C.~Di Bartolo, R.~Gambini and J.~Pullin,
  Phys.\ Rev.\  D {\bf 74}, 124012 (2006)
  [arXiv:gr-qc/0610023].

\bibitem{master}
  T.~Thiemann,
  Class.\ Quant.\ Grav.\  {\bf 23}, 2249 (2006)
  [arXiv:gr-qc/0510011] and references therein.

\bibitem{horava}
  P.~Horava,
  Phys.\ Rev.\  D {\bf 79}, 084008 (2009)
  [arXiv:0901.3775 [hep-th]].

\bibitem{collins}
  J.~Collins, A.~Perez, D.~Sudarsky, L.~Urrutia and H.~Vucetich,
  Phys.\ Rev.\ Lett.\  {\bf 93}, 191301 (2004)
  [arXiv:gr-qc/0403053].


\bibitem{Husain:2010gb}
  V.~Husain and A.~Kreienbuehl,
 Phys.\ Rev.\  D {\bf 81}, 084043 (2010)
  [arXiv:1002.0138 [gr-qc]].

\bibitem{HoHuSe}
  G.~M.~Hossain, V.~Husain and S.~S.~Seahra,
  arXiv:1007.5500 [gr-qc].


\bibitem{Laddha:2010hp}
  A.~Laddha and M.~Varadarajan,
  arXiv:1001.3505 [gr-qc].

\bibitem{thiemannscalar}
  T.~Thiemann,
  Class.\ Quant.\ Grav.\  {\bf 15}, 1487 (1998)
  [arXiv:gr-qc/9705021];
  T.~Thiemann,
  Class.\ Quant.\ Grav.\  {\bf 15}, 1281 (1998)
  [arXiv:gr-qc/9705019].

\bibitem{harmonicoscillator}
  A.~Ashtekar, S.~Fairhurst and J.~L.~Willis,
  Class.\ Quant.\ Grav.\  {\bf 20}, 1031 (2003)
  [arXiv:gr-qc/0207106].


\bibitem{dewitt} 
  B.~S.~DeWitt,
  Phys.\ Rev.\  {\bf 162}, 1195-1239 (1967).

\bibitem{time}
  R.~Gambini, R.~A.~Porto, J.~Pullin and S.~Torterolo,
  Phys.\ Rev.\  D {\bf 79}, 041501 (2009)
  [arXiv:0809.4235 [gr-qc]].

\bibitem{obregon}
  R.~Gambini, R.~Porto, J.~Pullin,
  Gen.\ Rel.\ Grav.\  {\bf 39}, 1143-1156 (2007).

\bibitem{ng}
F. K\'arolhy\'azy, A. Frenkel, B. Luk\'acs in ``Quantum concepts in
space and time'' R. Penrose and C. Isham, editors, Oxford University
Press, Oxford (1986); Y.~J.~Ng and H.~van Dam, Annals N.\ Y.\ Acad.\
Sci.\ {\bf 755}, 579 (1995) [arXiv:hep-th/9406110]; Mod.\ Phys.\
Lett.\ A {\bf 9}, 335 (1994); G.~Amelino-Camelia, 
Measurability Of Space-Time Distances In The Semiclassical
3415 (1994) [arXiv:gr-qc/9603014]; S. Lloyd, J. Ng, Scientific
American, November (2004).


\bibitem{rovelli}
  C.~Rovelli and S.~Speziale,
  Phys.\ Rev.\  D {\bf 67}, 064019 (2003)
  [arXiv:gr-qc/0205108].

\end{thebibliography}
\end{document}